\documentclass[a4paper,11pt]{article}
\usepackage{graphicx,amssymb,bm,latexsym,color,epsf}
\makeatletter
\@addtoreset{equation}{section}

\makeatother
\newcommand{\be}{\begin{equation}}
\newcommand{\ee}{\end{equation}}
\newcommand{\bea}{\begin{eqnarray}}
\newcommand{\eea}{\end{eqnarray}}
\newcommand{\vs}[1]{\vspace{#1 mm}}
\newcommand{\hs}[1]{\hspace{#1 mm}}
\renewcommand{\a}{\alpha}
\renewcommand{\b}{\beta}
\renewcommand{\c}{\gamma}
\renewcommand{\d}{\delta}

\newcommand{\s}{\sigma}

\newcommand{\vp}{\varphi}

\newcommand{\pa}{\partial}

\newcommand{\nn}{\nonumber\\}
\newcommand{\p}[1]{(\ref{#1})}

\newcommand{\bR}{\bar R}
\newcommand{\bg}{\bar g}

\newcommand{\Det}{{\rm Det}}

\def\lich{{\Delta_L}}

\begin{document}

\begin{titlepage}

\renewcommand{\thefootnote}{\fnsymbol{footnote}}
\begin{flushright}
KU-TP 071 \\
\today
\end{flushright}

\vs{10}
\begin{center}
{\Large\bf Quantum Equivalence of $f(R)$ Gravity and Scalar-tensor Theories in the Jordan and Einstein Frames}
\vs{15}

{\large
Nobuyoshi Ohta%$^{b,c,}$
\footnote{e-mail address: ohtan@phys.kindai.ac.jp}
} \\
\vs{10}

{\em Department of Physics, Kindai University,
Higashi-Osaka, Osaka 577-8502, Japan}

and

{\em Maskawa Institute for Science and Culture,
Kyoto Sangyo University, Kyoto 603-8555, Japan}

\vs{15}
%%%%%%%%%%%%%%%%%%%%%%%%%%%%%%%%
{\bf Abstract}
\end{center}

The $f(R)$ gravity and scalar-tensor theory are known to be equivalent at the classical level.
We study if this equivalence is valid at the quantum level.
There are two descriptions of the scalar-tensor theory in the Jordan and Einstein frames.
It is shown that these three formulations of the theories give the same determinant or effective action
on shell, and thus they are equivalent at the quantum one-loop level on shell
in arbitrary dimensions.
We also compute the one-loop divergence in $f(R)$ gravity on an Einstein space.

%%%%%%%%%%%%%%%%%%%%%%%%%%%%%%%%

\end{titlepage}
\newpage
\setcounter{page}{2}
\renewcommand{\thefootnote}{\arabic{footnote}}
\setcounter{footnote}{0}

%%%%%%%%%%%%%%%%%%%%%%%%%%%%%%%%
\section{Introduction}
%%%%%%%%%%%%%%%%%%%%%%%%%%%%%%%%

It is always of great interest to consider various modifications of Einstein gravity for phenomenological
applications. Among others, what is called $f(R)$ gravity attracts much attention especially
in the context of inflationary scenario in cosmology.
For early attempts, see \cite{HN}-\cite{CCT} and \cite{SF,DT} for reviews.
This class of theories has a nice feature that even though the theory involves higher derivatives,
there is no ghost introduced. In addition to the massless spin two graviton, the theory involves an additional
scalar degree of freedom.
The simplest way to see this is to rewrite the theory using scalar field coupled to the Einstein theory,
leading to scalar-tensor theory.
It has been known for long time that the equivalence is valid at the classical level on shell
(see, for example~\cite{Schmidt,Maeda,SF}), but it has not been much discussed at the quantum level.

One possible way to understand quantum properties of gravity is the asymptotic safety
scenario~\cite{W}-\cite{Reuter2012}. The idea is that the theory of quantum gravity is searched for within
a large class of theories, and one single theory is chosen by the condition that it corresponds to
a fixed point of the renormalization group flow. The use of functional renormalization group equation has given
a considerable evidence in support of the existence of a nontrivial fixed point.
The asymptotic safety scenario is discussed for $f(R)$ gravity in \cite{cpr1}-\cite{FO},
and for scalar-tensor theory in \cite{NO}-\cite{Henz}.

In fact, Benedetti and Guarnieri considered the problem of the equivalence by using the functional renormalization
group approach~\cite{BG}. They rewrite the $f(R)$ gravity in the scalar-tensor theory in the form without kinetic
term for the scalar, and then introduced kinetic term for the scalar with constant coefficient $\omega$.
After deriving functional renormalization group equation in Feynmann and Landau gauges, they try to
find fixed points in these gauges, in particular for $\omega=0$ case. They found that the results
disagree with each other, and argued that this is an evidence against the equivalence to $f(R)$ theory.
However this might be just a gauge artifact.

While our work is in progress, a paper appeared in which one-loop divergence in $f(R)$ gravity on arbitrary background
has been computed in a specific gauge~\cite{RS1}.
Based on this result, it is argued that $f(R)$ gravity and classically equivalent scalar-tensor theory are also
equivalent on shell at the quantum level~\cite{RS2}. It would be interesting to further check the equivalence at
the level of effective potential and/or functional renormalization group satisfied by the effective average action.
As the equivalence holds on shell classically, it is expected that the equivalence only holds on shell
also at the quantum level.
It is a general belief that the effective potential is gauge independent on shell~\cite{benedettionshell},
so it is expected that the equivalence holds on shell in any gauges. We also expect that the result is
independent of the parametrization of the metric.
In this paper, we discuss this problem by studying the effective actions or equivalent determinants
obtained in the path integral formulation in arbitrary dimensions.

There are two equivalent (at the classical level) formulations of scalar-tensor theories related by conformal
transformation. They are known as scalar-tensor theories in the Jordan and Einstein frames.
There is still ongoing debate about the quantum equivalence of these theories in the different
frames~\cite{MS}-\cite{KP}.
We also study the relation in this paper.
We find that all these formulations are equivalent on shell at the quantum level.

This paper is organized as follows. In sect.~2, we first review how the $f(R)$ theory is rewritten
into a theory of scalar field coupled to the Einstein theory in the Jordan frame. Then we make conformal
transformation to map the theory in the Einstein frame.
In sect.~3, we start studying the effective actions in these theories in the background field formalism.
In subsect.~3.1, we derive Hessian for the metric fluctuation $h_{\mu\nu}$ in $f(R)$ gravity on Einstein background,
which is assumed throughout this paper.
Using the exponential parametrization of the metric which has nice feature that it gives results rather close to
on-shell~\cite{nink,falls,pv,opv,opv2,opp,opp2}, we calculate the determinant with general linear gauge having
two gauge parameters $\a$ and $\b$. We show that the resulting effective action or determinant after path integral
does not depend on the gauge parameters (if we make partial gauge fixing $h_\mu^\mu=0$).
For completeness, we also give the one-loop divergent part of the effective action and the resulting functional
renormalization group equation. The first agrees with the recent calculation~\cite{RS1}.
In subsect.~3.2, we repeat the calculation in the scalar-tensor theory in the Jordan frame.
The Hessian has matrix structure but after taking the determinant, we find that the result precisely agrees
with that in the $f(R)$ theory if we use field equations for the background.
In subsect.~3.3, we go on to the scalar-tensor theory in the Einstein frame, and find that the resulting
determinant is different off shell but becomes the same on shell. This is to be expected because classically
the theory is equivalent only on shell.
We take these facts as evidence of the quantum equivalence of these theories.
In sect.~4, we give conclusions and discussions.
In the appendix, we give a formula of conformal transformation.

\section{Classical Equivalence}

Let us consider the Euclidean theory
\bea
S_{f}=\int d^d x \sqrt{g} f(R),
\label{frg}
\eea
where $g= \mbox{det}(g_{\mu\nu})$.
Classically it is known that this theory is equivalent to a scalar field $\phi$ coupled to
the Einstein gravity.

To move to such a formulation, let us first consider the theory
\bea
S=\int d^d x \sqrt{g} \Big[f'(\chi)(R-\chi) + f(\chi)\Big],
\label{form1}
\eea
where $\chi$ is a new scalar field.
If we take the variation with respect to $\chi$ and $g^{\mu\nu}$, we get
\bea
&&f''(\chi)(R-\chi)=0, \nonumber  \\
&&f'(\chi)R_{\mu\nu}-\frac{1}{2}[f'(\chi)(R-\chi) + f(\chi)] g_{\mu\nu}
- \nabla_\mu\nabla_\nu f'(\chi) + g_{\mu\nu} \nabla^2 f'(\chi) =0.
\label{fe1}
\eea
We assume that $f''(\chi)\neq 0$, and then we get $\chi=R$. Substituting this into \p{form1}
or the second equation in \p{fe1}, we find that it reduces to
\bea
f'(R)R_{\mu\nu}-\frac{1}{2} f(R) g_{\mu\nu}  - \nabla_\mu\nabla_\nu f'(R) + g_{\mu\nu} \nabla^2 f'(R) =0,
\label{master}
\eea
which is nothing but the field equation obtained from the action~\p{frg}.
Thus the theory~\p{form1} is classically equivalent to \p{frg}.

We can now rewrite the theory further using another scalar field $\phi$. We set
\bea
Z_N \phi = -f'(\chi),
\label{rel}
\eea
where $Z_N=\frac{1}{16\pi G}$ with $G$ being the Newton constant.
It should be understood that we solve \p{rel} for the field $\chi$ in terms of $\phi$.
Then Eq.~\p{form1} takes the form
\bea
S=\int d^d x \sqrt{g} \Big[Z_N \phi \{ \chi(\phi)-R \} + f(\chi(\phi))\Big].
\label{form2}
\eea
The field equations following from this action are
\bea
\d\phi:&&\hs{-5} Z_N (\chi(\phi)-R +\phi \chi'(\phi))+f'(\chi(\phi))\chi'(\phi)=0, \nn
\d g^{\mu\nu}: &&\hs{-5} Z_N (- \phi R_{\mu\nu}+ \nabla_\mu\nabla_\nu\phi -g_{\mu\nu}\nabla^2 \phi)
-\frac{1}{2}\Big[ Z_N \phi \{ \chi(\phi)-R \}+f(\chi(\phi))\Big]g_{\mu\nu} = 0.
\label{fe2}
\eea
Here and in what follows, the primes should be understood as differentiations with respect to the arguments,
so $f'(\chi)=\frac{df(\chi)}{d\chi}$ and $\chi'(\phi)=\frac{d\chi(\phi)}{d\phi}$,
and they should not be confused.
Using \p{rel} in the first equation, we find $\chi(\phi)=R$. Together with \p{rel} again,
the second equation in \p{fe2} then recovers \p{master}.
So this theory is also classically equivalent to \p{frg}.

We define a potential by
\bea
V(\phi) = Z_N \phi \chi(\phi)+f(\chi(\phi)).
\eea
Using \p{rel}, the derivatives of the potential is found to be
\bea
V'(\phi) &=& Z_N \chi(\phi), \nn
V''(\phi) &=& Z_N \chi'(\phi).
\label{jp}
\eea
We also have
\bea
\chi'(\phi) = -\frac{Z_N}{f''(\chi)}.
\label{est}
\eea
The action~\p{form2} is what is known as a theory of scalar field coupled to gravity in the Jordan frame.
We refer to this theory as scalar-tensor theory in the Jordan frame.

We can go to the Einstein frame by setting
\bea
g_{\mu\nu} = \phi^{-2/(d-2)} \tilde g_{\mu\nu}.
\label{ng}
\eea
With the help of the formula in the appendix, the action~\p{form2} is transformed into
\bea
S=\int d^d x \sqrt{\tilde g}\, \left[ Z_N \Big\{ - \tilde R + \frac{d-1}{d-2}\frac{(\pa_\mu\phi)^2}{\phi^2}
+ \phi^{-2/(d-2)} \chi(\phi)\Big\} + \phi^{-d/(d-2)} f(\chi(\phi)) \right].
\label{form3}
\eea
We change the scalar kinetic term by setting
\bea
\ln\phi=\sqrt{\frac{d-2}{d-1}}\ \vp,
\label{phivp}
\eea
and get
\bea
S_{EF}=\int d^d x \sqrt{\tilde g}\, \left[ Z_N \Big\{ - \tilde R + (\pa_\mu \vp)^2
 + e^{- \frac{2}{\sqrt{(d-1)(d-2)}}\vp} \chi(\vp)\Big\}
 + e^{- \frac{d}{\sqrt{(d-1)(d-2)}}\vp} f(\chi(\vp)) \right].
\label{form3}
\eea
We also refer to this theory as scalar-tensor theory in the Einstein frame.
This should be again equivalent to \p{frg}.
Thus we have two equivalent formulations of the theory~\p{frg} at the classical level.
Note that this equivalence is valid on shell, i. e. when we use the field equations.
The question that we would like to address is
whether these descriptions are also equivalent at quantum level.
We expect that the equivalence is also valid only on shell.

From \p{phivp}, we have
\bea
\vp = \sqrt{\frac{d-1}{d-2}} \ln \left(\frac{-f'(\chi)}{Z_N}\right) \,,
\eea
and
\bea
\chi'(\vp) = \sqrt{\frac{d-2}{d-1}} \frac{f'(\chi)}{f''(\chi)} \,,  \quad
\chi''(\vp)= \frac{d-2}{d-1} \frac{f'(\chi)}{f''(\chi)} \frac{f''(\chi)^2 -f'(\chi) f'''(\chi)}{f''(\chi)^2}
\quad etc.  
\eea
We can get all the equations for $\chi^{(n)}(\vp)$ in terms of $f$ and its derivatives. 
Then if we define the potential by
\bea
U(\vp) = e^{- \frac{d}{\sqrt{(d-1)(d-2)}}\vp}\Big\{
 Z_N e^{\sqrt{\frac{d-2}{d-1}}\ \vp} \chi(\vp) + f(\chi(\vp))\Big\},
\eea
we can express the condition of a minimum of the potential in terms of $f(\chi)$:
\bea
U'(\vp) = \left(\frac{-f'(\chi)}{Z_N}\right)^{- \frac{d}{d-2}}
 \frac{ 2\chi f'(\chi) - d\,f(\chi)}{\sqrt{(d-2)(d-1)}}.
\eea
In addition we also have that
\bea
U(\vp) = \left(\frac{-f'(\chi)}{Z_N}\right)^{- \frac{d}{d-2}}\left(f(\chi) -  \chi f'(\chi)\right)\,.
\eea

The Einstein equations for $\tilde{g}_{\mu\nu}$ is
\bea
-Z_N \tilde R_{\mu\nu} -\frac{1}{2} \tilde g_{\mu\nu} \left[-Z_N \tilde R +Z_N(\pa_\rho \vp)^2+U(\vp)\right]
+Z_N \pa_\mu\vp\pa_\nu\vp=0.
\eea
For constant backgrounds, this gives
\bea
Z_N \tilde{R} = \frac{d}{d-2} U(\vp).
\eea
On the other hand, by the transformation~\p{ng}, we also have that
\bea
\tilde{R}  = \left(\frac{-f'(\chi)}{Z_N}\right)^{\frac{-2}{d-2}} R
\eea
At the minimum of the potential, we have
\bea
U(\vp)_{min} = \left(\frac{-f'(\chi)}{Z_N}\right)^{- \frac{2}{d-2}}\frac{d-2}{d} Z_N \chi.
\eea
and we get the Einstein equation
\bea
\left(\frac{-f'(\chi)}{Z_N}\right)^{\frac{-2}{d-2}} R = \left(\frac{-f'(\chi)}{Z_N}\right)^{\frac{-2}{d-2}}  \chi,
\eea
or
\bea
R=\chi.
\eea

\section{Quantum equivalence}

In order to discuss quantum theory,
we use the background field method and expand the metric and the scalar fields as
\newcommand{\tp}{\tilde\phi}
\newcommand{\tvp}{\tilde\vp}
\bea
g_{\mu\nu}= \bg_{\mu\rho} ( e^h )^\rho{}_\nu\ ,\qquad
\phi=\bar\phi + \tp, \qquad
\vp=\bar\vp + \tvp.
\label{para}
\eea
We will consider constant background $\bar\phi$ and $\bar\vp$.
Note that we use the exponential parametrization for the metric. This is because this parametrization
has various virtues like least gauge-dependence.

For the one-loop calculation, we have to derive the Hessian.
%For simplicity, tildes on the fluctuation fields will be suppressed in the following.
Henceforth we assume that the background space is an Einstein space with
\bea
\bR_{\mu\nu} = \frac{\bR}{d} g_{\mu\nu},~~~
\bR = \mbox{const.}
\label{ein}
\eea

\subsection{$f(R)$ gravity}

For the $f(R)$ gravity, we have the quadratic term~\cite{opv2}
\bea
I^{(2)}_{f(R)}\hs{-2}&=&\hs{-2}
-\frac{1}{4} f'(\bR) h_{\mu\nu}^{TT} \Big(\lich_{2}-\frac{2}{d}\bR \Big) h^{TT\, \mu\nu}
 \nn && \hs{-4}
+ \frac{d-1}{4d} \s \Big[ \frac{2(d-1)}{d} f''(\bR) \Big(\lich_0 -\frac{\bR}{d-1}\Big)
+ \frac{d-2}{d}f'(\bR)\Big] \lich_0^2 \Big( \lich_0-\frac{\bR}{d-1} \Big) \s
 \nn && \hs{-5}
+\frac{1}{4} h \Big[ \frac{2(d-1)^2}{d^2} f''(\bR) \Big(\lich_0-\frac{\bR}{d-1} \Big)^2
+\frac{(d-1)(d-2)}{d^2} f'(\bR) \Big(\lich_0-\frac{2}{d-2}\bR\Big)
+ \frac{1}{2} f(\bR) \Big] h
 \nn && \hs{-5}
+ \frac{d-1}{2d} h \Big[ \frac{2(d-1)}{d} f''(\bR) \Big(\lich_0-\frac{\bR}{d-1} \Big)
+\frac{d-2}{d} f'(\bR) \Big]
 \lich_0 \Big(\lich_0-\frac{\bR}{d-1}\Big) \s,
\label{hessianfr}
\eea
where we have suppressed the overall factor $\sqrt{\bg}$ and
$\lich_2$, $\lich_1$ and $\lich_0$
are the Lichnerowicz Laplacians on the symmetric tensor, vector and scalar respectively,
defined by
\bea
\lich_{2} T_{\mu\nu} &=& -\nabla^2 T_{\mu\nu} +R_\mu{}^\rho T_{\rho\nu}
+ R_\nu{}^\rho T_{\mu\rho} -R_{\mu\rho\nu\s} T^{\rho\s} -R_{\mu\rho\nu\s} T^{\s\rho}, \nn
\lich_{1} V_\mu &=& -\nabla^2 V_\mu + R_\mu{}^\rho V_\rho, \nn
\lich_{0} S &=& -\nabla^2 S.
\eea
We have also used the York decomposition defined by
\bea
h_{\mu\nu} = h^{TT}_{\mu\nu} + \nabla_\mu\xi_\nu + \nabla_\nu\xi_\mu +
\nabla_\mu \nabla_\nu \s -\frac{1}{d} \bg_{\mu\nu} \nabla^2 \s +
\frac{1}{d} \bg_{\mu\nu} h,
\label{york}
\eea
where
\bea
\nabla_\mu h^{TT}_{\mu\nu} = \bg^{\mu\nu} h^{TT}_{\mu\nu}
= \nabla_\mu \xi^\mu=0.
\eea
The above formula agrees with \cite{ms}.
In terms of $s=\lich_0 \s+h$, Eq.~\p{hessianfr} is put into
\bea
I^{(2)}_{f(R)} \hs{-2}&=&\hs{-2}
-\frac{1}{4} f'(\bR) h_{\mu\nu}^{TT} \left(\lich_2-\frac{2}{d}\bR \right) h^{TT\, \mu\nu}
 \nn && \hs{-5}
+\, \frac{(d-1)^2}{2d^2} s \left[  f''(\bR) \lich_0 + \frac{(d-2)f'(\bR)-2\bR f''(\bR)}{2(d-1)} \right]
\left( \lich_0-\frac{\bR}{d-1} \right) s \nn
&& +\, \frac{df(\bR)-2\bR f'(\bR)}{8d} h^2
\label{hessianfr2}
\eea

We then consider the gauge fixing term
\bea
\label{gfaction}
S_{GF}=\frac{1}{2\alpha}\int d^d x \sqrt{\bg}\,\bg^{\mu\nu}F_\mu F_\nu,
\eea
with
\bea
\label{gf}
F_\mu=\nabla_\rho h^\rho{}_\mu-\frac{\b+1}{d}\nabla_\mu h\ .
\eea
Following \cite{benedettionshell}, it is convenient to reparametrize the scalar sector in terms of the
gauge-invariant variable $s$ and a new degree of freedom $u$ defined as
\bea
\label{chi}
s=h+\lich_{0} \s,~~~
u=\frac{[(d-1)\lich_{0}-\bR]\s+\b h}{(d-1-\b)\lich_{0} -\bR}.
\eea
The gauge fixing action then becomes
\bea
\label{gf2}
S_{GF}=\frac{1}{2\alpha}\int dx\sqrt{\bg}
\left[\xi_\mu\left(\lich_{1}-\frac{2\bR}{d}\right)^2\xi^{\mu}
+\frac{(d-1-\beta)^2}{d^2}
u \lich_{0} \left(\lich_{0}-\frac{\bR}{d-1-\beta}\right)^2 u \right].
\eea
On shell, the last term in \p{hessianfr2} is zero,
so the quadratic part of the action is written entirely
in terms of the physical degrees of freedom $h^{TT}$ and $s$,
and the gauge fixing entirely in terms of the gauge degrees of freedom
$\xi$ and $u$.

The ghost action for this gauge fixing contains a non-minimal operator
\bea
S_{gh}=\int dx\sqrt{\bg}\bar C^\mu\left(
\delta_\mu^\nu\nabla^2
+\left(1-2\frac{\beta+1}{d}\right)\nabla_\mu\nabla^\nu+\bR_\mu{}^\nu\right)C_\nu .
\eea
Upon decomposing the ghost into transverse and longitudinal parts
\bea
C_\nu=C^T_\nu+\nabla_\nu C^L
=C^T_\nu+\nabla_\nu\frac{1}{\sqrt{\lich_{0}}}C'^L ,
\eea
and the same for $\bar C$, the ghost action splits in two terms
\bea
\label{ghostaction}
S_{gh}=\int dx\sqrt{\bg}
\left[
-\bar C^{T\mu}\left(\lich_{1}-\frac{2\bR}{d}\right)C^T_\mu
-2\frac{d-1-\beta}{d}
\bar C'^L\left(\lich_{0}-\frac{\bR}{d-1-\beta}\right)C'^L\right].
\eea

Now if we make a partial gauge fixing to set $h=0$, which can be done by sending $\b \to \infty$,
we get the following one-loop determinants:
\bea
\Det\left[ \Delta_{L2}-\frac{2}{d}\bR \right]^{-1/2}
\Det \left[ \left(  f''(\bR) \lich_0 + \frac{(d-2)f'(\bR)-2\bR f''(\bR)}{2(d-1)} \right)
 \left( \Delta_{L0}-\frac{\bR}{d-1} \right)\right]^{-1/2}, \nn
% \nn && \times
\eea
from \p{hessianfr2},
\bea
\Det \left[\Delta_{L1} -\frac{2\bR}{d} \right]^{-1} \Det \left[ \Delta_{L0}\right]^{-1/2}
\Det \left[ \Delta_{L0}-\frac{\bR}{d-1-\b} \right]^{-1},
\eea
from the gauge fixing term~\p{gf2}, and
\bea
\Det\left[\Delta_{L1} -\frac{2\bR}{d} \right]
\Det\left[ \Delta_{L0}-\frac{\bR}{d-1-\b} \right],
\eea
from the ghost terms~\p{ghostaction}.
The York decomposition has Jacobian
\bea
\label{yorkj}
{\Det} \Big(\lich_{1}-\frac{2}{d}\bR \Big)^{1/2}
\Det [\lich_0]^{1/2}
{\Det} \Big(\lich_{0}-\frac{\bR}{d-1} \Big)^{1/2}
\eea
whereas the subsequent transformation $(\sigma,h)\to(s,u)$ has unit Jacobian.
We see many of these cancel and we are left with
\bea
\frac{
{\Det} \Big[\lich_{1}-\frac{2}{d}\bR \Big]^{1/2}}
{{\Det} \Big[\lich_2-\frac{2\bR}{d} \Big]^{1/2}
{\Det} \Big[\lich_0 + \frac{(d-2)f'(R)-2 \bR f''(\bR)}{2(d-1)f''(\bR) } \Big]^{1/2}} \ .
\label{oneloopea}
\eea
As observed in \cite{opv2} and confirmed in \cite{opp2}, this result does not depend on the gauge parameters
$\a$ and $\b$, and the result is close to on-shell once the partial gauge choice $h=0$ is made.
This is the advantage of the exponential parametrization~\cite{nink,falls,pv,opv,opv2,opp,opp2}.
The above determinant is what governs the quantum theory at the one-loop level, in particular effective action.

Given the above result, we can evaluate the effective action which is related to the partition function by
$Z(\bg)=e^{-\Gamma(\bg)}$. Neglecting field-independent terms, we find
\bea
\Gamma(\bg) &=&
\frac{1}{2}\log\Det\left(\lich_2-\frac{2\bR}{d}\right)
-\frac{1}{2}\log\Det\left(\lich_1-\frac{2\bR}{d}\right) \nn
&& +\frac{1}{2}\log\Det\left(\lich_0-\frac{\bR}{d-1}+\frac{(d-2)f'(\bR)}{2(d-1)f''(\bR)}\right)\ .
\label{divea}
\eea
The divergent part of the effective action
can be computed by standard heat kernel methods \cite{perbook}.
On an Einstein background in four dimensions, the logarithmically divergent part is
\bea
\Gamma_{log}(\bg)= \frac{1}{12(4\pi)^2}
\int d^4x\,\sqrt{\bg}
\log\left(\frac{\Lambda^2}{\mu^2}\right)\left(
-\frac{71}{10}\bR_{\mu\nu\rho\sigma}^2
+\frac{433}{120}\bR^2 - \frac{f'(\bR)^2}{3 f''(\bR)^2}+\frac{\bR f'(\bR)}{f''(\bR)}
\right)
\,,
\label{gammaabc}
\eea
where $\Lambda$ stands for a cutoff and we introduced a reference mass scale $\mu$.
On shell, we have $f'(\bR)=\frac{2f(\bR)}{\bR}$, and $\bR_{\mu\nu\rho\sigma}^2=\frac{\bR^2}{6}$
for maximally symmetric space, this reduces to
\bea
\Gamma_{log}(\bg)= \frac{1}{24(4\pi)^2}
\int d^4x\,\sqrt{\bg}
\log\left(\frac{\Lambda^2}{\mu^2}\right)\left(
\frac{97}{20}\bR^2 - \frac{8f(\bR)^2}{3 \bR^2 f''(\bR)^2}+\frac{4 f(\bR)}{f''(\bR)}
\right)
\,,
\eea
in agreement with \cite{RS1}.

The flow equation for the $f(R)$ theory on the 4-sphere was derived in \cite{opv,opv2}.
The result using spectral sum is %\tc{red}{This should be modified probably.}
\bea
&& \hs{-10}
32 \pi^2 (\dot \Phi -2r \Phi'+4\Phi) \nn
&& \hs{-10} = \frac{d_1}{6+(6 \alpha+1) r}-\frac{d_2 \left(2 r \Phi ''-2 \Phi'
-\dot{\Phi }' \right)}{\Phi '}+\frac{d_3 \left(\dot{\Phi }''-2 r \Phi'''\right)
+d_4 \Phi ''}{(3+(3 \beta -1) r) \Phi ''+\Phi '}+\frac{d_5}{4+(4 \gamma -1) r} \,,~~~
\label{erge_spectral}
\eea
where
\bea
d_1&=&\frac{5 [6+(6\a-1)r]\, [12+(12\a-1)r]}{12} \,,\nn
d_2&=& \frac{5 [6+(6\a-1)r]\, [3+(3\a-2)r]}{108}
\,,\nonumber\\
d_3&=&\frac{ [2+(2\b+3)r]\, [3+(3\b-1)r]\, [6+(6\b-5)r]}{72}\,,\nn
d_4&=&\frac{[2+(2 \beta -1)r]\, [12+(12 \beta +11)r]}{8}\,,\nn
d_5&=&\frac{-72-18 r (1+8 \gamma )+r^2 \left(19-18 \gamma -72 \gamma ^2\right)}{6}\,.
\label{coe_spectral}
\eea
and we have used the dimensionless quantities $r =\bR k^{-2}$ and
$\Phi(r) = k^{-4} f(\bR)$, and $\a, \b$ and $\c$ are the parameters of endomorphism, not to be confused with
the gauge parameters.

\subsection{Scalar-tensor theory in the Jordan frame}

Next we discuss the one-loop determinant in the scalar-tensor theory in the Jordan frame.

We find from \p{jp} that the quadratic terms in the fluctuations in the scalar field are
\bea
\sqrt{g} V(\phi) \simeq \sqrt{\bg} \left[ \frac{Z_N}{2} \chi'(\bar\phi) (\tilde\phi^2 + h\tilde\phi)
+\frac{1}{8}V(\bar\phi) h^2 \right]
\eea
Together with the contribution from the rest of the terms, we find, using the York decomposition~\p{york},
\bea
I^{(2)} &=& - Z_N \bar\phi\left[ -\frac{1}{4} h^{TT}_{\mu\nu} \left(\lich_2-\frac{2\bR}{d} \right) h^{TT\mu\nu}
+ \frac{(d-2)(d-2)}{4d^2} s \left(\lich_0 -\frac{\bR}{d-1} \right) s + \frac{d-2}{8d} \bR h^2 \right] \nn
&& -Z_N \left[ \frac{\bR}{2}h\tilde\phi +\frac{d-1}{d} \tilde\phi \left(\lich_0 -\frac{\bR}{d-1} \right) s \right]
+\frac{Z_N}{2} \chi'(\bar\phi) (\tilde\phi^2 + h\tilde\phi)
+\frac{1}{8}V(\bar\phi) h^2 .
\eea
We employ the same gauge fixing as in the preceding subsection.
Here we also make the partial gauge fixing $h=0$. Then we again find that the quadratic part of the action
is written entirely in terms of the physical degrees of freedom $h^{TT}, s$ and $\tilde\phi$
and the gauge fixing entirely in terms of the gauge degrees of freedom $\xi$ and $u$.
The one-loop determinant from the scalar sector $(s,\tilde\phi)$ is
\bea
&& \Det \left(\begin{array}{cc}
\frac{(d-1)(d-2)}{2d^2} Z_N \bar\phi\left(\lich_0-\frac{\bR}{d-1}\right) & 
- \frac{d-1}{d} Z_N \bar\phi\left(\lich_0-\frac{\bR}{d-1}\right) \\
- \frac{d-1}{d} Z_N \bar\phi\left(\lich_0-\frac{\bR}{d-1}\right) &
Z_N \chi'(\bar\phi)
\end{array} \right)^{-1/2}
\nn &&
= \Det \left[ \frac{(d-1)^2}{d^2} Z_N^2 \left(\lich_0-\frac{\bR}{d-1}\right)
\left(\lich_0-\frac{\bR}{d-1}+\frac{d-2}{d-1} \bar\phi\chi'(\bar\phi)\right)\right]^{-1/2}
\eea
It follows from \p{rel} and \p{est} that
\bea
\phi \chi'(\phi) = \frac{f'(\chi)}{f''(\chi)}.
\eea
So, writing out the resulting whole one-loop determinant, we get
\bea
\frac{
{\Det} \Big[\lich_{1}-\frac{2}{d}\bR \Big]^{1/2}}
{{\Det} \Big[\lich_2-\frac{2\bR}{d} \Big]^{1/2}
{\Det} \Big[\lich_0 + \frac{(d-2)f'(\bar\chi)}{2(d-1)f''(\bar\chi)}-\frac{\bR}{d-1} \Big]^{1/2}} \ .
\label{oneloopeaJ}
\eea
If we use $\bar \chi=\bR$, this precisely agrees with \p{oneloopea}, the result for the $f(R)$ theory.
Thus we conclude that this formulation by scalar-tensor theory is equivalent to the original $f(R)$ theory
at the quantum (at least) one-loop level on shell.

\subsection{Scalar-tensor theory in the Einstein frame}

Next we discuss the Hessian and one-loop determinant in the scalar-tensor theory in the Einstein frame.

If we make one more step in the discussion in sect.~2, we find that
the second derivative of the potential is given by
\bea
U''(\vp) = \left(\frac{-f'(\chi)}{Z_N}\right)^{- \frac{2}{d-2}} Z_N \frac{(d-2) f'-2 \chi f''}{(d-1) f''}
- \frac{d}{\sqrt{(d-2)(d-1)} }U'(\vp),
\eea
so the Hessian for the fluctuation $\tilde \vp$ is
\bea
I^{(2)}_{\vp\vp} = Z_N\left( \lich_0 + %\left(\frac{-f'(\chi)}{Z_N}\right)^{-\frac{2}{d-2}}
\frac{(d-2) f'-2 \chi f''}{2(d-1)f''} \right) - \frac{d}{2 \sqrt{(d-2)(d-1)} }U'(\bar\vp).
\eea
Note that if we exploit the equations of motion for the background $U'(\bar\vp)=0$ and $\bar\chi=\bR$, we find that
this is proportional to that of the field $s$ in \p{hessianfr2}:
\bea
S^{(2)}_{ss} \propto  \left. \frac{f''}{Z_N %\left(\frac{-f'(\chi)}{Z_N}\right)^{- \frac{2}{d-2}}
}
S^{(2)}_{\vp\vp} \right|_{ \mbox{\scriptsize on-shell}}.
\eea
There are also some mixing terms between graviton $h$ and the scalar $\vp$, but these drop out in the gauge $h=0$
in the exponential parametrization.

The rest of the theory is the usual Einstein theory.
The Hessian for this theory can be obtained from the result in the preceding subsection by setting $f(R)=-Z_N R$.
We thus find
\bea
I^{(2)}_{E}\hs{-2}&=&\hs{-2}
-Z_N \left[ - \frac{1}{4} h_{\mu\nu}^{TT} \Big(\lich_{2}-\frac{2}{d}\bR \Big) h^{TT\, \mu\nu}
+\frac{(d-1)(d-2)}{4d^2} s \left( \lich_0- \frac{\bR}{d-1}\right) s
+ \frac{d-2}{8d} \bR h^2 \right].\nn
\label{hessianei}
\eea
The gauge fixings can be taken as in the preceding subsections.
With the partial gauge fixing $h=0$, we again find the separation of the physical degrees of freedom
and the gauge fixing terms.
It is now straightforward to derive the one-loop determinant
\bea
\frac{
{\Det} \Big[\lich_{1}-\frac{2}{d}\bR \Big]^{1/2}}
{{\Det} \Big[\lich_2-\frac{2\bR}{d} \Big]^{1/2}
{\Det} \Big[Z_N\left( \lich_0 + %\left(\frac{-f'(\chi)}{Z_N}\right)^{-\frac{2}{d-2}}
\frac{(d-2) f'-2 \chi f''}{2(d-1)f''} \right) - \frac{d}{2 \sqrt{(d-2)(d-1)} }U'(\vp) \Big]^{1/2}} \ .
\label{oneloopeaE}
\eea
As noted above, on shell, this is equivalent to the result of $f(R)$ gravity.

However before concluding that the theory is equivalent at the quantum level, we have to take the Jacobian from
the transformation~\p{phivp} into account. The path integral would produce divergences in the form $\d(0)$ times
the volume from this change of variable. However such terms would affect the power-law divergence coefficients,
which are subject to regularization scheme. The coefficients of logarithmic divergence are not affected and are
universal. It is true that we have to take into account the difference in the definition of the scales in different
frames, but that can be easily incorporated since the form of the effective action are the same.
One may worry that the conformal transformation introduces the change of path integral measure and hence leads to
a trace anomaly. However it will be taken into account in a form of the logarithmic ultraviolet (UV) cutoff dependence
when the determinant is regularized with UV cutoff, which is dependent on the conformal transformation in terms of
the scalar field~\cite{HKN}. We will discuss this problem of the difference in the scale in the next section.
All the results for one-loop divergence, effective action and flow equation can be derived from these determinants.
The scale dependence in different frames would introduce a formal difference in the resulting functional
renormalization group equations, but should not affect the physical results.
We thus conclude that the theory is also equivalent to the $f(R)$ theory at quantum level on shell.

\section{Conclusions and discussions}

In this paper, we first summarized the relation of the $f(R)$ theory and the reformulations
of the theory in the form of a scalar field coupled to the Einstein theory in the Jordan and Einstein frames,
and then calculated determinants, which correspond to the effective actions, after path integral over
the fluctuation fields in the background field formalism.
It turns out that all three formulations give the same determinant on shell.
If we evaluate the determinant with a suitable cut off, this gives the divergences in the theory.
After renormalization, this produces an effective action. We have also given the one-loop divergent term
in the effective action. One could also try to derive the functional
renormalization group equation by introducing suitable cutoff function.
The fact that the determinant, from which these are all derived, are the same is a strong evidence
that these theories are equivalent at the quantum level, at least at one loop.

One possible caveat is that the transformation into the Einstein frame involves conformal transformation.
This transformation would produce $\delta(0)$ type divergence, which could be removed by a local counterterm.
However this also produce difference in the scales in different frames. In a given frame, short distance cutoff $\ell$
may be defined by
\bea
\ell^2 = g_{\mu\nu} \Delta x^\mu \Delta x^\nu.
\eea
When the metric is transformed as \p{ng}, the cutoff lengths in the Jordan and Einstein frames are related by
\bea
\ell_J^2 = g_{\mu\nu}^J \Delta x^\mu \Delta x^\nu = (\bar\phi)^{-2/(d-2)} g_{\mu\nu}^E \Delta x^\mu \Delta x^\nu
= (\bar\phi)^{-2/(d-2)} \ell_E^2.
\eea
The UV cutoff is then related by
\bea
\Lambda_J^2 = (\bar\phi)^{2/(d-2)} \Lambda_E^2.
\eea
This would result in slightly different looking functional renormalization group equations in the two frames
due to the different cutoffs.
If this difference is dealt with suitably, the physical result should not depend on the difference
because the effective action is the same.
For related discussions, see \cite{HKN}.

To summarize, we have found strong evidence that the $f(R)$ theory and the scalar-tensor theories are
equivalent on shell in arbitrary dimensions. As a byproduct, we also find evidence that the theories
in the Jordan and Einstein frames are equivalent.
Note that our discussions are based on the Einstein space with the curvature~\p{ein}.
It would be interesting to try to extend our result to more general spacetime.

%%%%%%%%%%%%%%%%%
\section*{Acknowledgment}
%%%%%%%%%%%%%%%%%
We would like to thank Kevin Falls for valuable discussions at the early stage of this work,
and Roberto Percacci for valuable comments.
This work was supported in part by the Grant-in-Aid for
Scientific Research Fund of the Japan Society for the Promotion of Science (C) No. 16K05331.

\appendix

\section{Conformal transformation}

We give a formula relevant in the text here.
Under the transformation
\bea
g_{\mu\nu}= e^{-2\rho} \tilde g_{\mu\nu},
\eea
the Einstein term changes as
\bea
\sqrt{g} R = \sqrt{\tilde g}\, e^{(2-d)\rho} \left[\tilde R+2(d-1)\nabla_\mu^2 \rho-(d-1)(d-2) (\pa_\mu\rho)^2\right].
\eea
Note that the contraction is made on the right hand side by $\tilde g$.

%%%%%%%%%%%%%%%%%%%%%%%%%%%%%%%%%

\end{document}